\documentclass[pra,twocolumn,preprintnumbers,amsmath,amssymb,nofootinbib,showpacs]{revtex4-1}
\usepackage{graphicx}
\usepackage{amssymb}
\usepackage{mathrsfs}
\usepackage{color}

\newcommand{\beq}{\begin{equation}}
\newcommand{\eeq}{\end{equation}}

\begin{document}
\preprint{FPAU0-15/16}

\title{Chiral $p\pm ip$ superfluid on a sphere}

\author{Sergej Moroz$^{1,2}$, Carlos Hoyos$^3$ and Leo Radzihovsky$^{1,2}$}
\affiliation{$^1$Department of Physics, University of Colorado, Boulder, Colorado 80309, USA \\
                $^2$Center for Theory of Quantum Matter, University of Colorado, Boulder, Colorado 80309, USA \\
                $^3$Department of Physics, Universidad de Oviedo, Avda. Calvo Sotelo 18, 33007, Oviedo, Spain}

\begin{abstract}
We consider a spinless fermionic $p\pm ip$ superfluid living on a two-dimensional sphere. Using superfluid hydrodynamics we show that the ground state necessarily exhibits topological defects: either a pair of elementary vortices or a domain wall between $p\pm ip$ phases. In the topologically nontrivial BCS phase we identify the chiral fermion modes localized on the topological defects and compute their low-energy spectrum. 
\end{abstract}


\maketitle

\section{Introduction} \label{intro}

Ever since the discovery of superfluidity in $^3\text{He}$ \cite{Osheroff1972,Osheroff1972b}, \emph{chiral} fermionic superfluids stepped into spotlight of low-temperature physics \cite{vollhardt,volovik1992exotic,volovikbook}. Of particular interest are \emph{two-dimensional} fully gapped chiral superfluids with the condensate
\beq \label{gapflat}
\hat \Delta=|\Delta_0| (p_x \pm i p_y),
\eeq
where $p_i=-i \partial_i$ and the sign defines the chirality.
In addition to the particle number symmetry, these exotic superfluids break spontaneously continuous rotational symmetry as well as discrete parity and time-reversal. Spin polarized two-dimensional fermions with a short-range attractive interaction, investigated before in the cold atom experiment \cite{Gunter2005}, is the simplest model that gives rise to such a superfluid. Its realization using p-wave Feshbach resonances has been extensively investigated
theoretically and demonstrating the topological and other phase transitions \cite{Gurarie2005,Gurarie20072}. Moreover, modern advances in nanofabrication allowed to create and study chiral superfluids in thin films of $^3\text{He}$ \cite{Levitin2013,Levitin2014}. Also the Moore-Read $\nu=5/2$ quantum Hall state is a $p_x+ip_y$ superfluid of composite fermions \cite{Read2000}.

It is instructive to consider a chiral superfluid living on a \emph{curved} two-dimensional surface. In this case the chiral condensate  that appears in the pairing Hamiltonian density can be written as
\beq \label{gapcurved}
\hat \Delta=|\Delta_0| e^{-2i\theta} (e^1 \pm i e^2)^i p_i,
\eeq
where $e^{aj}$ is a vielbein, i.e. a pair ($a=1,2$) of orthonormal vectors defined at every point of the surface. In addition, we introduced the Goldstone phase $\theta$, which is not necessarily uniform in the ground state.\footnote{Indeed, since the vielbein is not unique and can be locally rotated in the internal vielbein space without affecting the metric, our construction possesses an internal \emph{gauge redundancy}. The Goldstone phase transforms by a shift under internal vielbein rotations \cite{Hoyos2013} and thus can be non-uniform in the ground state.}  See Appendix \ref{App0} for the details on how Eq. \eqref{gapflat} is generalized to the form \eqref{gapcurved}.

\begin{figure}[ht]
\begin{center}
\includegraphics[height=0.25\textwidth]{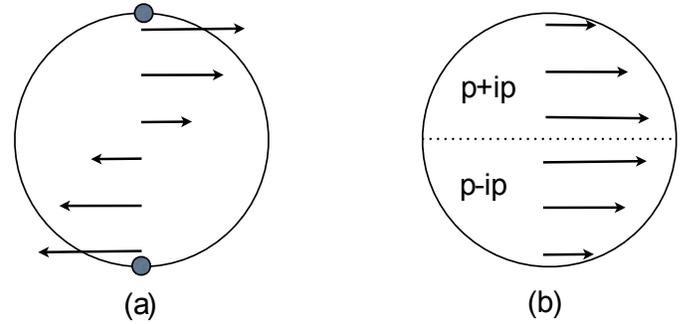}
\caption{Ground state candidates for a chiral superfluid on a sphere: (a) the vortex pair solution for the $p-ip$ pairing, (b) the domain wall solution between
$p+ip$ and $p-ip$ superfluids. The velocity field sketched schematically with arrows.}\label{fig1}
\end{center}
\end{figure}

In this paper, in particular, we investigate the structure of the ground state of a $p\pm ip$ superfluid living on a surface of a two-dimensional sphere. Contrary to the case of a conventional s-wave superfluid or a chiral superfluid on a flat substrate, the ground state of a chiral superfluid necessarily supports \emph{topological defects}. Mathematically, this follows from the Poincar\'e-Hopf (``hairy ball'') theorem that asserts that one can not define on a sphere a vielbein vector field without critical points. Using superfluid hydrodynamics, in Sec. \ref{CLL} we construct two candidates for the ground state shown in Fig. \ref{fig1}.

\begin{figure}[ht]
\begin{center}
\includegraphics[height=0.30\textwidth]{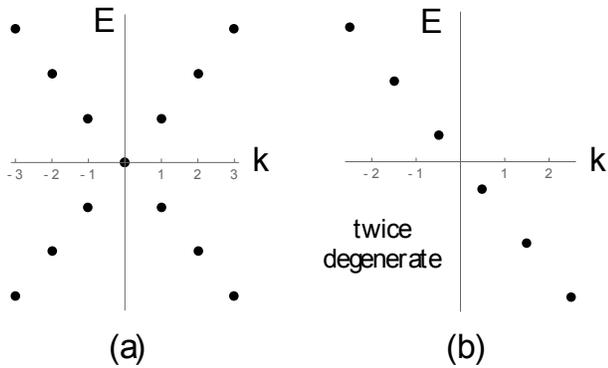}
\caption{Energy of chiral fermion modes as a function of angular momentum for (a) the vortex pair solution, (b) the domain wall solution with a strong repulsive potential localized around the equator which prohibits hybridization of the two chiral modes. }\label{fig2}
\end{center}
\end{figure}

It is well-known that in the topological BCS phase of a chiral superfluid, topological defects (vortices and domain walls) and edges bind \emph{chiral fermion} modes \cite{Volovik1999, Read2000, Stone2004, volovikbook, Sauls2011, Alicea2012}. In Sec. \ref{CFM} we solve the Bogoliubov-de Gennes (BdG) equation 
for the two ground state candidates from Fig. \ref{fig1} and determine the low-energy spectrum of the chiral fermion modes. The result is illustrated in Fig. \ref{fig2}.

Finally, we mention several previous studies that are related to our work.
In the late 1970s three-dimensional $^3\text{He-A}$ in a spherical container was investigated extensively, see \cite{ trickey1977, vollhardt}.
The physics of a \emph{vector} (and also tensor) order parameter on a curved two-dimensional surface has also been investigated in soft condensed matter. In particular, we refer to \cite{Park1992} (and references therein), where a closed deformable surface of genus zero was considered. In addition,
Read and Green argued in \cite{Read2000} that vortices will appear in the ground state of a two-dimensional $p\pm ip$ paired state defined on a sphere.
In more detail a $p\pm ip$ superconductor on a sphere was later investigated in \cite{Kraus2008, Kraus2009} (see also \cite{Moller2011}). In these papers, however, a $U(1)$ \emph{magnetic monopole} was introduced at the center of the sphere, which compensated the effect of the spherical geometry and guaranteed that the number of vortices and antivortices is the same.  In the present paper, we do not introduce the magnetic monopole and thus concentrate on the  physics resulting purely from the spherical geometry. We would also like to mention a recent paper \cite{Schi2015}, where a \emph{two-body chiral} problem on a two-dimensional sphere was solved.

\section{Ground state: vortex pair vs domain wall} \label{CLL}
In this section we investigate the nature of the chiral superfluid ground state on a sphere by using the low-energy effective theory developed in \cite{Hoyos2013, Moroz2014a} (see also \cite{Stone2004} for a precursor). At zero temperature a chiral spinless superfluid has one gapless Goldstone mode in the energy spectrum that governs the low-energy and long-wavelength dynamics. The superfluid velocity of a chiral superfluid is given by \cite{Hoyos2013}
\beq \label{vel}
v_i=-\partial_i \theta+s \omega_i+A_i,
\eeq
where $\theta$ is the Goldstone phase. In contrast to conventional s-wave superfluids \cite{Turner2010}, the remarkable property of a \emph{chiral} superfluid is that its velocity depends on the spin connection $\omega_i$ with the parameter $s=\pm 1/2$ for the $p\pm i p$ pairing. On a generic two-dimensional surface with a metric $g_{ij}=e^a_i e^a_j$ the spin connection is defined by 
$
\omega_i\equiv \frac 1 2 \epsilon^{ab}e^{aj} \nabla_i e^b_j,
$
where $\nabla_i$ is a covariant derivative and $\epsilon^{ab}$ is the antisymmetric Levi-Civita symbol. Despite the fermions being electrically neutral in our problem, we introduced here an external $U(1)$ gauge potential $A_\mu$. The corresponding $U(1)$ electric and magnetic fields can be switched on by introducing a gravitational field and by rotating the fermions, respectively.

It follows from Eq. \eqref{vel} that vortices in chiral superfluids are sourced not only by rotation ($U(1)$ magnetic field), but also by the Gaussian curvature. This is a direct superfluid analogue of the ``shift'' \cite{Wen1992} that has been recently studied extensively in quantum Hall physics. The ``shift'' was also introduced and computed for relativistic chiral superfluids \cite{Golkar2014,Golkar:2015dya}.

Consider now a unit sphere parametrized with spherical coordinates, the polar angle $\zeta\in(0,\pi)$ and the azimuthal angle $\phi\in(0,2\pi)$ with the metric
$g_{ij}=\text{diag}(1,\sin^2\zeta)$
 and the corresponding orthonormal vielbein vectors
$e^{\zeta i}=(1,0)^\text{T}$ and $e^{\phi i}=(0, 1/\sin\zeta)^\text{T}$. In these coordinates the spin connection is
$
\omega_i=(0, -\cos\zeta).
$
giving rise to a constant Gaussian curvature
$
K=\varepsilon^{ij} \partial_i \omega_j=1,
$
 where $\varepsilon^{ij}\equiv g^{-1/2} \epsilon^{ij}$.
The poles are \emph{singular} points of spherical coordinates and a proper treatment of the geometry of these points is given in  Appendix \ref{AppA}.

Curiously, the spin connection on the sphere is identical to the gauge potential of a $U(1)$ magnetic monopole of charge $q=+1$ placed at the center of the sphere \cite{Preskill1984}. As a result, by adding a $U(1)$ magnetic monopole of charge $q=-s$, we can completely compensate the effect of curvature and have a ground state with $\mathbf{v}=0$ and no topological defects. This has been done in the previous study of a $p\pm ip$ superconductor on a sphere \cite{Kraus2008, Kraus2009,Moller2011}, where a vortex-antivortex excitation above the ground state has been considered. Since it is not obvious how one can create experimentally a $U(1)$ monopole for neutral superfluids, we will not introduce it in in this paper.

\subsection{Vortex pair solution}
Consider a simple ansatz $\theta=0$ and $A_i=0$ resulting in the velocity field on a unit sphere
\beq
v_\zeta=0, \qquad
v_\phi=-s\cos \zeta.
\eeq
 At the equator ($\zeta=\pi/2$) the velocity vanishes, as it changes direction as one goes from the northern to the southern hemisphere (see Fig. \ref{fig1}). One can check that $\nabla_i v^i=0$ and $\varepsilon^{i j} \partial_i v_j=s$. The magnitude of the velocity field diverges at the poles, where we have a pair of (anti)vortices for $s=\pm 1/2$. Indeed, the vortex winding number is given by the circulation integral evaluated on an infinitesimal loop
\beq
\text{Winding number}=\frac 1 \pi \oint v=-2s.
\eeq

So far we have just guessed the form of the velocity. It turns out, however, that this stationary field satisfies the conservation equations of superfluid hydrodynamics
\beq \label{matcon}
\nabla_i J^i=0,
\eeq
\beq \label{momcon}
\nabla_i T^{ij}=0,
\eeq
where $J^i=-g^{-1/2} \delta S/\delta A_i$ and $T^{ij}=2g^{-1/2} \delta S/\delta g_{ij}$ is the $U(1)$ current and stress tensor, respectively. For a chiral superfluid they were computed in \cite{Hoyos2013}. In particular, one can check that Eq. \eqref{matcon} and Eq. \eqref{momcon} projected on the velocity field are automatically satisfied provided the superfluid density $\rho$ is only a function of $\zeta$. Based on symmetry we expect $\rho(\zeta)$ to be an even function around the equator. The precise from of the $\zeta$-dependence of the superfluid density $\rho$ is fixed by the scalar equation
$
e^{\zeta}_j\nabla_i T^{ij}=0.
$

\subsection{Domain wall between $p\pm ip$ phases}
Is there a solution on a sphere with no vortices? We start from the $p\pm ip$ superfluid that has a nontrivial winding of the Goldstone phase around the poles, i.e.,
$
\theta=- \text{sgn}(\pi/2 - \zeta) s \phi
$ and
$ A_\mu=0
$.
The resulting velocity field is
$
v_\zeta=0
$,
$
v_\phi=s(\pm 1-\cos \zeta)
$,
where the upper and lower sign should be taken in the northern and southern hemisphere, respectively. 
Note that the velocity field \emph{vanishes} at the poles and thus there are no vortices there.  At the equator, however, the azimuthal component of the velocity field is not continuous and undergoes a jump
$
\Delta v_\phi=2s.
$
This discontinuity can not appear in a physical solution and is resolved as follows: The superfluid spontaneously chooses the condensates of opposite chiralities in the northern and southern hemisphere ($s^{\text{N}}=-s^{\text{S}}$). The resulting velocity field on a unit sphere
\beq \label{dws}
v_\zeta=0, \qquad
v_\phi=s^{\text{N}}(1\mp \cos \zeta)
\eeq
has no vortices and is continuous at the equator (see Fig. \ref{fig1}). The phases of opposite chiralities are separated by a \emph{domain wall}.
It is straightforward to check that the velocity profile \eqref{dws} is a solution of the conservation equations \eqref{matcon} and \eqref{momcon}.
\subsection{Ground state}
At this point it is natural to ask what scenario from Fig. \ref{fig1} will be actually realized in the ground state. To answer this question, one has to evaluate the total energy of the superfluid for both scenarios and choose the one with a smaller energy. The result depends on the \emph{equation of state}, i.e. the internal energy $\epsilon(\rho)$. Indeed, on a general two-dimensional surface with the metric $g_{ij}$ the energy of a (chiral) superfluid is given by \cite{Moroz2014a}
\beq \label{en}
E=\int d^2 x \sqrt{g} \Big[ \frac {\rho v_i v_j g^{ij}}{2}+\epsilon(\rho)\Big].
\eeq
For the vortex pair solution, it will generically contain the core contribution that scales  $\sim \ln \big(\mathcal{R}/\xi_{\text{core}} \big)$, where $\mathcal{R}$ is the radius of the sphere and $\xi_{\text{core}}$ is the size of the vortex core that is fixed by the equation of state. On the other hand, the domain wall solution in addition to the energy \eqref{en} will contain the domain wall gradient energy\footnote{Since the energy of the domain wall is the largest, when it is located at the equator, one should worry about the stability of the the domain wall. This issue is discussed in Appendix \ref{AppB}.}
$ 
E_{\text{DW}}\sim \frac{|\Delta_0|^2 \mathcal{R}}{\xi_{\text{DW}}},
$
where $\xi_{\text{DW}}$ is the domain wall thickness, fixed by the equation of state and the structure of the interface. Intuitively, in the limit $\mathcal{R}\to \infty$ the domain wall solution is energetically unfavored with respect to the vortex pair solution. For a finite $\mathcal{R}$, however, the nature of the ground state depends on the equation of state and possibly either of the scenarios can be realized. One can thus speculate that a \emph{transition} between the two scenarios can be found by tuning a parameter such as the radius of the sphere $\mathcal{R}$.

Note that the scenarios shown in Fig. \ref{fig1} are the \emph{minimal} solutions in the sense that the topology of the sphere also allows to add an arbitrary number of vortex-antivortex pairs. Intuitively, these states should have higher energies as compared to the ones we discussed. This, however, should be carefully checked after the equation of state is fixed.

\section{Chiral fermion modes} \label{CFM}

As long as a chiral $p\pm ip$ superfluid is in the topological BCS phase, topological defects bind \emph{chiral fermion} modes on their surface \cite{Volovik1999, Read2000, Stone2004,Gurarie2007, volovikbook,Sauls2011} (see also the review \cite{Alicea2012} and references therein). In this section we investigate the properties of these modes for the two ground state candidates shown in Fig. \ref{fig1}.

\subsection{Index theorem}
The index theorem (see Sec. 22.1 in \cite{volovikbook}) relates the algebraic number\footnote{The algebraic number is the number of right-movers minus the number of left-movers.} $\nu$ of chiral modes localized at an interface between topologically distinct phases of a chiral superfluid to the change $\Delta C$ of the bulk Chern number
\beq
\nu=\Delta C.
\eeq
Based on this theorem we anticipate for
\begin{itemize}
\item Vortex pair state: Two chiral modes each localized on the North and South pole, respectively.\footnote{More precisely, on a sphere the two modes will \emph{hybridize} and thus the two resulting states will be localized on both poles simultaneously. If the radius of the sphere is large compared to the spatial extent of the chiral mode, the hybridization will have a small effect on the energy spectrum. \label{fthyb}} This follows from viewing a vortex in a $p\pm ip$ superfluid as an interface between the trivial vacuum core phase with $C=0$ and the topological outer BCS phase with $C=\pm 1$.
\item Domain wall state: Two co-propagating chiral modes localized at the equator because the domain wall separates $p+ip$ and $p-ip$ BCS phases with the Chern numbers $C=1$ and $C=-1$, respectively. 
\end{itemize}

\subsection{BdG equation on a sphere}
The index theorem guarantees the presence of chiral fermion modes in the ground state of a $p\pm ip$ superfluid in the BCS regime on a sphere. In order to find their energy spectrum,  we solve the BdG equation 
\beq
\mathcal H F=E F,
\eeq
where $F=(u, v)^\text{T}$ and the BdG Hamiltonian
\beq
\mathcal H=\Big ( \begin{tabular}{ cc}
  $\epsilon(p)$ & $\hat \Delta$ \\
  $\hat \Delta^\dagger$ & $-\epsilon(p)$  \\
\end{tabular}
 \Big)
\eeq
with $\epsilon(p)=g^{ij} p_i p_j/(2m)-\mu$ and $p_i=-i\nabla_i$. The fermions are spinless, so they do not couple directly to the spin connection. Since the form of the order parameter $\hat \Delta$ is different for the two scenarios from Fig. \ref{fig1}, we will now discuss the two cases separately.

\subsubsection{Vortex pair state} \label{vps}
Consider a unit sphere parametrized by spherical coordinates. 
With details relegated to Appendix \ref{AppC}, it can be shown that for the $p\pm ip$ superfluid with a pair of elementary (anti)vortices at the poles the order parameter \eqref{gapcurved} reduces to
\beq \label{opvp}
\begin{split}
\hat \Delta&=-i |\Delta_0| \Big(\partial_\zeta \pm \frac{i}{\sin \zeta}\partial_\phi \Big).
\end{split}
\eeq
Importantly, the presence of the (anti)vortex pair ensures that $\hat\Delta$ does not contain a nontrivial winding phase of the angle $\phi$.

First, we factorize the solution into a highly oscillating spherical harmonic $Y^k_l$ and a slowly changing spinor $\tilde F$, i.e.,
\beq
F=Y^{k}_{l_F} \tilde F,
\eeq
where  the Fermi angular momentum $l_F\gg 1$ was defined by
$
\mu=l_F(l_F+1)/(2m)\approx l_F^2/(2m)
$
and $-l_F\le k\le l_F$ is an \emph{integer}. In the following we consider the regime $|k|\ll l_F$.
The BdG equation transforms into
\beq
 \tilde {\mathcal  H} \tilde F=E \tilde F
\eeq
with the transformed BdG Hamiltonian to lowest order in the derivative (Andreev approximation)
\beq \label{andham}
\begin{split}
\tilde {\mathcal H}=&-\frac {\tau_3} m \Big(g^k_{l_F} \partial_\zeta+\frac {ik}{\sin^2\zeta} \partial_\phi \Big) \\
  &-i \tau_1|\Delta_0| g^k_{l_F} \mp \tau_2 |\Delta_0| \frac k {\sin\zeta},
\end{split}
\eeq
where we introduced  the function $g^k_l\equiv \partial_\zeta Y^k_l/ Y^k_l$. The explicit form of this function is not important, but we will use later that $g^k_{l_F}$ scales with $l_F$, i.e., $g^k_{l_F}\sim l_F$.
Consider first the case $k=0$, that reduces the Hamiltonian to
\beq
\tilde {\mathcal H_0}=g^k_{l_F} \Big( -\frac {\tau_3} m \partial_\zeta -i \tau_1|\Delta_0| \Big).
\eeq
We look for zero-energy solutions of this Hamiltonian, i.e., solve
\beq
\Big(-i \tau_3 \partial_\zeta+\tau_1 m |\Delta_0|\Big) \tilde F=0.
\eeq

There are two orthogonal \emph{Majorana} solutions each localized at the North and South pole, respectively
\beq \label{wfN}
\tilde F_N(\zeta)=\Big ( \begin{tabular}{ c}
   $1$ \\
   $-i$  \\
\end{tabular}
 \Big) \underbrace{\exp\Big(-m \int_0^\zeta dx \,  |\Delta_0|(x)\Big)}_{\psi^N_0(\zeta)},
\eeq
\beq \label{wfS}
\tilde F_S(\zeta)=\Big ( \begin{tabular}{ c}
   $1$ \\
   $i$  \\
\end{tabular}
 \Big) \underbrace{\exp\Big(-m \int_\zeta^\pi dx \,  |\Delta_0|(x)\Big)}_{\psi^S_0(\zeta)}.
\eeq
In the following we will assume that $|\Delta_0|$ is an even function around the equator and thus $\psi^N_0(\zeta)=\psi^S_0(\pi-\zeta)\equiv \psi_0(\zeta)$.

For $k\ne 0$ we will treat the remaining part of the Hamiltonian
\beq
\Delta \tilde {\mathcal H}=\mp \tau_2 |\Delta_0| \frac k {\sin\zeta}-\frac{ik \tau_3}{m \sin^2\zeta}\partial_\phi
\eeq
using degenerate perturbation theory which is well justified for $|k|\ll l_F$  because $\tilde {\mathcal H_0}\sim l_F$, while $\Delta \tilde {\mathcal H} \sim k$.
Using the wave-functions \eqref{wfN}, \eqref{wfS} we find
\beq
\begin{split}
\Delta E&=\Big ( \begin{tabular}{cc}
   $\pm k \omega_0$ & $0$ \\
   $0$ & $\mp k \omega_0$\\
\end{tabular}
 \Big), \\
\omega_0&= \frac{\int d\zeta |\psi_0(\zeta)|^2 |\Delta_0|(\zeta)/\sin\zeta}{\int d\zeta |\psi_0(\zeta)|^2}.
\end{split}
\eeq
The two chiral modes cross zero energy with two opposite slopes as a function of the (integer) angular momentum quantum number $k$, which is schematically illustrated in Fig. \ref{fig2}. Importantly, the slope $\omega_0$ is finite because the gap $|\Delta_0|$ vanishes (generically linearly) in the cores of the vortices located at the poles.

As already mentioned in the footnote \ref{fthyb}, since the two (anti)vortices are separated by a finite distance on a sphere, we expect them to \emph{hybridize}. As a result, the two zero modes will mix and acquire a finite energy gap. This effect goes beyond the present approximation.

\subsubsection{Domain wall state}
Previous studies (for a summary see \cite{Samokhin2012} and references therein, see also \cite{Bouhon2013}) of a domain wall between $p\pm ip$ phases in flat space demonstrated that the energy spectrum of the chiral fermion states is \emph{nonuniversal}, i.e., it depends on the microscopic details of the model. One basic reason behind that is the \emph{hybridization} of the two co-propagating modes localized at the interface \cite{Kwon2004} resulting in accumulation of the unbroken charge \cite{Volovik2014a}.  The nonuniversal aspects of the problem can be circumvented by introducing a strong repulsive potential centered at the interface. This potential introduces a thin topologially trivial BEC phase inside the domain wall and effectively creates two edges along which two decoupled chiral modes propagate.\footnote{The original domain wall can be recovered by gradually switching off the repulsive barrier.} Here we will follow this route and study the two edge states on a unit sphere. We expect that the low-energy spectrum of this problem is \emph{universal}.

The strong repulsive potential at the equator slices the sphere in half and thus it is sufficient to investigate only the edge of one (say southern) hemisphere. By symmetry of Fig \ref{fig1} (b) the physics of the other hemisphere edge state is the same. It is demonstrated in Appendix \ref{AppC} that the order parameter of a $p-ip$ superfluid in the southern hemisphere is given in spherical coordinates by
\beq \label{opdw}
\begin{split}
\hat \Delta&=-i |\Delta_0| e^{i\phi} \Big(\partial_\zeta - \frac{i}{\sin \zeta}\partial_\phi \Big).
\end{split}
\eeq
Here the winding phase factor is nontrivial. This phase factor and the gap profile $|\Delta_0|$ that does not vanish at the poles, make the calculation of the kind performed in Sec. \ref{vps} technically difficult.

Nevertheless, we will argue below that for a \emph{smooth edge}\footnote{The edge will be called smooth if the chemical potential varies sufficiently slowly such that the superfluid density goes to zero over a length scale that is much larger than the superfluid coherence length $\xi\sim 1/|\Delta_0|$ \cite{Huang2015}. On a sphere of radius $\mathcal{R}$, the superfluid density  $\rho\sim 1/\mathcal{R}^2$ which gives rise to the condition $\mathcal{R} |\Delta_0|\gg 1$.} the present problem on a sphere can be reduced to the flat space edge problem on a disc that has already been solved \cite{Read2000} (see also the review \cite{Alicea2012} and references therein). To this end we perform the \emph{stereographic projection} that maps the southern hemisphere to a unit disc. In stereo-polar coordinates $(R,\Phi)$ the metric $g_{ij}=\chi^2 g_{ij}^{\text{p}}$, where $\chi=2/(1+R^2)$ is the Riemann conformal factor and $g_{ij}^{\text{p}}=\text{diag}(1,R^2)$ is the flat space metric expressed in polar coordinates. In these coordinates the Laplace operator $\nabla^2 f=\frac{1}{\sqrt{g}}\partial_i (g^{ij} \sqrt{g}\partial_j f)=\chi^{-2} \nabla^2_{\text{p}} f$, where $\nabla^2_{\text{p}}$ is the Laplace operator in flat space expressed in polar coordinates. The $p-ip$ order parameter in the southern hemisphere now reads
\beq
\begin{split}
\hat \Delta&=-i  \chi^{-1} |\Delta_0| e^{i\Phi} \Big(\partial_R-\frac{i}{R}\partial_\Phi \Big)=\chi^{-1} \hat \Delta_{\text{p}}.
\end{split}
\eeq
The BdG Hamiltonian can thus be written as
\beq
\mathcal H=\Big ( \begin{tabular}{ cc}
  $-\chi^{-2}\frac{\nabla^2_{\text{p}}}{2m}-\mu$ & $\chi^{-1}\hat \Delta_{\text{p}}$ \\
  $\chi^{-1} \hat \Delta^\dagger_{\text{p}}$ & $\chi^{-2}\frac{\nabla^2_{\text{p}}}{2m}+\mu$  \\
\end{tabular}
 \Big)
\eeq
which looks like a flat space Hamiltonian, but decorated with the Riemann conformal factors $\chi$. Notice, however, that for a smooth edge we can neglect the term quadratic in derivatives \cite{Read2000,Alicea2012}. In addition, we can absorb  the conformal factor into the definition of the gap, i.e., $|\Delta_0|\to \chi |\Delta_0| $. As a result, the problem becomes equivalent to the flat space superfluid on a disc with a smooth edge. In that case for the $p-ip$ pairing the low-energy spectrum of the chiral mode is known to be 
$
E=-|\Delta_0| k/ R_{\text{d}},
$
where $k$ is a \emph{half-odd integer}\footnote{It is the nontrivial winding phase in the order parameter that results in a \emph{half-odd integer} angular momentum quantum number $k$.} and $R_{\text{d}}$ is the radius of the disc.

As a final result, on a sphere of radius $\mathcal{R}$ we find two co-propagating modes localized at the domain wall (with a strong repulsive potential at the equator) with the \emph{twice degenerate} low-energy spectrum
\beq
E=-\frac {|\Delta_0|}{ \mathcal{R}} k,
\eeq
which is valid for $|\Delta_0| \mathcal{R}\gg 1$ and schematically illustrated in Fig. \ref{fig2} (b).

\section{Conclusions and outlook} \label{Conc}

We showed that topological defects necessarily appear in the ground state of a two-dimensional $p\pm ip$ superfluid confined to a sphere. Physically this happens because chiral Cooper pairs $\emph{rotate}$ and thus feel the Gaussian curvature as a kind of magnetic field. In this paper we identified the two candidates for the ground state that are illustrated in Fig. \ref{fig1}. In the topological BCS phase we also identified the fermion chiral modes localized on the defects and computed their low-energy spectrum (see Fig. \ref{fig2}), which generally is consistent with what happens in flat space. The basic reason behind this agreement is that, in contrast to Cooper pair, the elementary \emph{spinless} fermions do not couple to the spin connection and thus do not acquire a geometric Aharonov-Bohm phase on a sphere.   

It is straightforward to extend our findings to closed geometries of \emph{nonzero genus} $g$. Specifically, in this case the Gauss-Bonnet theorem suggests that a candidate for the ground $p\pm ip$ state must have the vortex number\footnote{The vortex number is the number of vortices minus the number of antivortices.} equal to $\mp \chi_E$, where the Euler characteristic $\chi_E=2-2g$. In particular, on a torus ($g=1$) the total vortex number is zero. It would be interesting to extend our work to curved geometries with \emph{boundaries}. 

In general our work suggests that for a $p\pm ip$ superfluid on a generic curved surface a $2\pi$ flux of the Gaussian curvature should give rise to a topological defect such as an (anti)vortex. We hope that further advances in $^3\text{He}$ experiments in nanofabricated geometries and cold atom experiments will make it possible to test our predictions and provide new directions for the extension of this work.

\section*{Acknowledgments:}
We acknowledge discussions with Victor Gurarie, Jaacov Kraus, Manfred Sigrist and Dam Thanh Son.
SM and LR acknowledge support by the NSF grants DMR-1001240, and by the Simons Investigator award from the Simons Foundation.
This work is partially supported by the Spanish grant MINECO-13-FPA2012-35043-C02-02. C.H is supported by the Ramon y Cajal fellowship RYC-2012-10370. 
\appendix
\section{Order parameter on arbitrary two-dimensional surface} \label{App0}

Consider first a many-body system of identical spin-polarized fermions $\psi$ living in flat space and interacting via the two-body potential $V(\mathbf{r})$. We will use Cartesian coordinates. The Cooper order parameter is the \emph{scalar} function
\beq \label{gapdef}
\Delta(\mathbf{r}_1, \mathbf{r}_2)=-V(\mathbf{r}_1-\mathbf{r}_2)\langle \psi (\mathbf{r}_1) \psi (\mathbf{r}_2) \rangle
\eeq
or equivalently
\beq
\Delta(\mathbf{R}, \mathbf{r})=-V(\mathbf{r})\langle \psi (\mathbf{R}+\mathbf{r}/2) \psi (\mathbf{R}-\mathbf{r}/2) \rangle,
\eeq
where we introduced the center-of-mass coordinate $\mathbf{R}=(\mathbf{r}_1+ \mathbf{r}_2)/2$ and the relative coordinate $\mathbf{r}=\mathbf{r}_1- \mathbf{r}_2$. It is often convenient to perform a Fourier transform $\mathbf{r}\to \mathbf{p}$. In the mean-field theory the order parameter acts on the conjugate of the fermion field as a matrix in position space
\beq \label{gene}
\hat \Delta \psi^* (\mathbf{r}_1)=\int d \mathbf{r}_2 \Delta(\mathbf{r}_1, \mathbf{r}_2) \psi^*(\mathbf{r}_2).
\eeq
We will now specialize to the order parameter that has \emph{chiral} symmetry in the relative space
\beq \label{ff}
\Delta(\mathbf{R}, \mathbf{p})=\Delta_{0}(\mathbf{R}) \mathbf{e}^{\pm}  \cdot \mathbf{p}.
\eeq
Here $\mathbf{e}^{\pm}=\mathbf{e}_x\pm i\mathbf{e}_y$ and $\mathbf{e}^x$ and $\mathbf{e}^y$ is a pair of \emph{constant} orthonormal vectors (vielbein) that point along $x$ and $y$ axis, respectively. Due to the \emph{quasi-local} nature of the order parameter, the integral in Eq. \eqref{gene} can be performed analytically. Specifically, we first Fourier transform $\mathbf{p}\to \mathbf{r}$ in Eq. \eqref{ff}, change the coordinates $\mathbf{r}, \mathbf{R}\to \mathbf{r}_1, \mathbf{r}_2$ and substitute the resulting expression into Eq. \eqref{gene}. After several integrations by parts we finally find\footnote{Note that this expression also holds, when the vielbein is \emph{not constant}, the situation to be discuss below.}
\beq \label{imp}
\hat \Delta \psi^* (\mathbf{r}_1)=\{ \hat p_i , e^{\pm \, i} \Delta_{0}(\mathbf{r}_1)\} \psi^*(\mathbf{r}_1),
\eeq
where $\hat p_i= -i {\partial}/{\partial r_1^i}$ and $\{\hat a,\hat b\}=(\hat a \hat b+\hat b \hat a)/2$. 

Since a vielbein is not unique, we will consider next the vielbein that depends on $\mathbf{R}$. Two-dimensional space is still flat and Cartesian coordinates are used. As a result, even in a uniform case we will need now to introduce the $\mathbf{R}$-dependent Goldstone phase $\theta$ by writing $\Delta_0(\mathbf{R})=|\Delta_0| e^{-2i \theta(\mathbf{R})}$ with $|\Delta_0|=\text{const}$. The chiral order parameter should be now written as
\beq
\Delta(\mathbf{R}, \mathbf{p})=\Delta_0(\mathbf{R})  \mathbf{e}^{\pm}(\mathbf{R})\cdot \mathbf{p}.
\eeq
After a chain of manipulations Eq. \eqref{imp} in this case can be put into a  compact form
\beq \label{gg}
\hat \Delta \psi^*=|\Delta_0| e^{-i \theta} \mathbf{e}^{\pm} \cdot \mathbf{\hat P} e^{-i\theta}\psi^*,
\eeq
where $\hat P_i \theta \equiv -i(\partial_i \theta-s \omega_i)$ and $\hat P_i \psi^*\equiv -i \partial_i \psi^*$. Here $s=\pm 1/2$ for the $p \pm ip$  condensate.

For an arbitrary (curved) manifold Eq. \eqref{gg} still holds. It can be derived from Eq. \eqref{imp} by using the identity
$\nabla_i e_j^b=-\omega_i \epsilon^{bc} e^c_j$. Up to a gauge, this identity fixes the vielbein field. As a result, the vielbein is covariantly constant under a covariant derivative that acts both on coordinate ($i,j,\dots$) and vielbein ($a,b,\dots$) indices. 

We are now in position to derive Eq. \eqref{gapcurved} that was stated in Sec. \ref{intro} without a proof. Indeed the pairing term of the mean-field Hamiltonian density
\beq
\begin{split}
\psi^* \hat \Delta \psi^*&= \psi^*|\Delta_0| e^{-i \theta} \mathbf{e}^{\pm} \cdot \mathbf{\hat P} e^{-i\theta}\psi^* \\
&=\psi^* \underbrace{|\Delta_0| e^{-2i \theta} \mathbf{e}^{\pm} \cdot \mathbf{\hat p}}_{\hat\Delta \, \text{from Eq. \eqref{gapcurved}}} \psi^*,
\end{split}
\eeq
where we used the local fermionic property $\psi^* \psi^*=0$ in order to put the expression into the final form.

\section{Geometry of poles} \label{AppA}
In spherical coordinates the poles are singular points of the coordinates. Indeed, we consider a small loop around a pole and calculate the circulation integral
$\oint \omega=\int \omega_\phi d\phi$. By Stokes theorem it should be proportional to the curvature flux penetrating the loop. Since a sphere is an orientable manifold, we can define a positive (counterclockwise) direction of the loop consistently. On the northern (southern) pole $\oint=\pm \int_{0}^{2\pi}$. In the limit of infinitesimally small loop around the North (South) pole we find
$
\oint \omega=-2\pi.
$
Although the curvature is finite everywhere on the sphere, the finite result for this integral makes it clear that the two poles are singular points of spherical coordinates. This problem can be resolved by gauge transforming to $\omega_\phi=\pm 1-\cos\zeta$ for the North (South) patch of the sphere, respectively. Now $K=1$ everywhere and thus the artificial curvature singularities disappear at the poles at the expense of multivaluedness of the spin connection $\omega_\phi$ in the overlap region. In this region the spin connections from the two patches are simply related by a gauge transformation \cite{Preskill1984}. 
\section{Stability of the domain wall at the equator} \label{AppB}
Imagine that the boundary between $p+ip$ and $p-ip$ phases is moved from the equator to the angle $\zeta_0\ne \pi/2$. The domain wall energy
$
E_{\text{DW}} \sim \frac{|\Delta_0|^2 \mathcal{R} \sin\zeta_0}{\xi_{\text{DW}}}
$
has a maximum at the equator.
On the other hand, one expects that the kinetic energy of the superfluid has a minimum at $\zeta_0=\pi/2$. For example, in the incompressible limit ($\rho=\text{const}$) one finds
\beq
\begin{split}
E_{\text{kin}}&=s^2 \pi \rho \Big[ \int_{0}^{\zeta_0} d\zeta \frac{(1-\cos \zeta)^2}{\sin\zeta}+\int_{\zeta_0}^{\pi} d\zeta \frac{(1+\cos \zeta)^2}{\sin\zeta} \Big] \\
&=s^2 \pi \rho \Big[-2+\log 16-4 \log \sin \zeta_0 \Big],
\end{split}
\eeq
which indeed has a minimum at $\zeta_0=\pi/2$. In general, the domain wall is stable at the equator if the total energy $E_{\text{tot}}$ satisfies
$
\partial^2_{\zeta_0}E_{\text{tot}}>0.
$
If the stability condition is violated, one expects that the domain wall will move away from the equator to some angle $\zeta_0=\pi/2\pm \Delta \zeta$ with $\Delta\zeta>0$.  The sign will be chosen spontaneously. One can determine $\Delta\zeta$ by minimizing $E_{\text{tot}} (\zeta_0)$ with respect to $\zeta_0$.

\section{Order parameter on a unit sphere} \label{AppC}
We start in flat space and consider a $p\pm ip$ superfluid with a vortex localized at the origin and having the vortex winding number $n$. The order parameter kernel in the $(\mathbf{R}, \mathbf{p})$ representation (see Eq. \eqref{ff} in Appendix \ref{App0}) is
\beq \label{firste}
\Delta(\mathbf{R},\mathbf{p})=e^{in\Phi} |\Delta_0| (p_x\pm i p_y),
\eeq
where the magnitude of the gap $|\Delta_0|$ is a function of $R$, $\Phi$ is the polar angle of the center of mass vector $\mathbf{R}$. Eq. \eqref{firste} can be expressed in the relative polar coordinates $(r,\phi)$ as
\beq \label{firste}
\Delta(\mathbf{R},\mathbf{p})=e^{in\Phi} |\Delta_0| e^{\pm i \phi} \underbrace{(p_r\pm \frac{i}{r} p_\phi)}_{\mathbf{e}^{\pm}_{\text{p}}\cdot \mathbf{p}}.
\eeq
Using now Appendix \ref{App0}, the order parameter operator can be written as
\beq
\begin{split}
\hat \Delta&=e^{i(n\pm 1)\phi} |\Delta_0|(r) \mathbf{e}^{\pm}_{\text{p}} \cdot \mathbf{p}, \\
&=-i e^{i(n\pm 1)\phi} |\Delta_0|(r) \big(\partial_r\pm \frac i r \partial_\phi \big).
\end{split}
\eeq
Incidentally, the order parameter agrees with the general form \eqref{gapcurved} with $\theta=-(n-1)\phi/2$. Importantly, we also find that in general the winding number of a vortex depends not only on the winding of the phase $\phi$, but also on the winding of the vielbein.

Since close to the poles spherical coordinates reduce to polar coordinates, it is now straightforward to write down the corresponding order parameter operator on a unit sphere. In particular, on the northern hemisphere we obtain for the $p\pm ip$ pairing
\beq \label{eqAA}
\begin{split}
\hat \Delta&=e^{i(n\pm 1)\phi} |\Delta_0|(\zeta) \mathbf{e}^{\pm}_{\text{sp}} \cdot \mathbf{p}, \\
&=-i e^{i(n\pm 1)\phi} |\Delta_0|(\zeta) \big(\partial_\zeta \pm \frac {i} {\sin \zeta} \partial_\phi \big),
\end{split}
\eeq
where $\mathbf{e}^{\pm}_{\text{sp}}$ is the chiral combination of the spherical vielbein vectors $\mathbf{e}^\zeta$ and $\mathbf{e}^\phi$, introduced in Sec. \ref{CLL}. Importantly, on a sphere the direction of the angular momentum of a $p\pm ip$ Cooper pair points in the \emph{opposite} directions on the North and South pole, respectively. This is why on the southern hemisphere the gap operator for the $p\pm ip$ pairing should be written as
\beq \label{eqBB}
\begin{split}
\hat \Delta&=e^{-i(n\pm 1)\phi} |\Delta_0|(\zeta) \mathbf{e}^{\pm}_{\text{sp}} \cdot \mathbf{p}, \\
&=-i e^{-i(n\pm 1)\phi} |\Delta_0|(\zeta) \big(\partial_\zeta \pm \frac {i} {\sin \zeta} \partial_\phi \big).
\end{split}
\eeq
 
Finally, we note that Eqs. \eqref{eqAA}, \eqref{eqBB} reduce to the order parameter operators \eqref{opvp} and \eqref{opdw} used in the main text for $n\pm1=0$ ((anti)vortex pair) and $n=0$ (no vortices), respectively.

\bibliography{library}

\end{document}